# Monk: Opportunistic Scheduling to Delay Horizontal Scaling

Marina Shimchenko[a] , Erik Österlund[b] , and Tobias Wrigstad[a]

a    Uppsala University, Sweden
b    Oracle, Sweden

**Abstract**    In modern server computing, efficient CPU resource usage is often traded for latency. Garbage collection is a key aspect of memory management in programming languages like Java, but it often competes with application threads for CPU time, leading to delays in processing requests and consequent increases in latency. This work explores if opportunistic scheduling in ZGC, a fully concurrent garbage collector (GC), can reduce application latency on middle-range CPU utilization, a topical deployment, and potentially delay horizontal scaling. We implemented an opportunistic scheduling that schedules GC threads during periods when CPU resources would otherwise be idle. This method prioritizes application threads over GC workers when it matters most, allowing the system to handle higher workloads without increasing latency. Our findings show that this technique can significantly improve performance in server applications. For example, in tests using the SPECjbb2015 benchmark, we observed up to a 15 % increase in the number of requests processed within the target 25 ms latency. Additionally, applications like Hazelcast showed a mean latency reduction of up to 40 % compared to ZGC without opportunistic scheduling. The feasibility and effectiveness of this approach were validated through empirical testing on two widely used benchmarks, showing that the method consistently improves performance under various workloads. This work is significant because it addresses a common bottleneck in server performance—how to manage GC without degrading application responsiveness. By improving how GC threads are scheduled, this research offers a pathway to more efficient resource usage, enabling higher performance and better scalability in server applications.

## The Art, Science, and Engineering of Programming

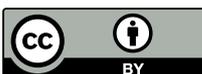

# Monk: Opportunistic Scheduling to Delay Horizontal Scaling

## 1 Introduction

OpenJDK is one of the most popular platforms for Java, estimated by Oracle to run 60 billion JVM instances [23]. Java is widely used for developing, for example, web applications [8], ranging from small websites to large-scale enterprise-level applications [28]. These applications serve various purposes, such as e-commerce platforms, content management systems, social media platforms, and online banking portals. Systems like these typically handle large volumes of concurrent incoming requests with varying frequencies, leading to varying CPU loads.

Driven by quality-of-service aspects, aka Server Level Agreements (SLAs) [11], server workloads often operate under a strict ceiling on the average CPU utilization [31, 34]. If an application exceeds this threshold, vendors tend to scale up (horizontal scaling) by migrating the application to another server to maintain acceptable latency levels [31]. For example, Amazon Kinesis Data Analytics [4] scales up automatically when the CPU utilization remains at 75 % or above for 15 minutes and scales down (decreases parallelism) when the CPU utilization remains below 10 % for six hours [3]. Alibaba Realtime Compute [2] has a similar policy. Shekhar Shashank et al. [26] explored how scaling affects latency-sensitive workloads. For their experiments, they chose a CPU utilization range of 50–70 %. They argue for this choice as they do not want the server to be underutilized or saturated. Whenever the utilization grows/reduces from the target range, they add or remove a core, respectively. We are happy to add a discussion to the paper that better motivates and explains why we are interested in this particular CPU utilization range.

To give an insight into how CPU utilization affects latency and why applications scale up after reaching a certain CPU load, let us consider a response curve of the SPECjbb2015 benchmark (see Section 4 for more details) that simulates an online store (Figure 1). The benchmark gradually increases the number of requests a machine can handle and measures the maximum response time, among other parameters. We identified equations that describe the data on low (up to 55 %), middle (55–80 %), and medium-high CPU loads (80–93 % and 93–100 %). The division is not precise, but it is good enough to showcase the differences in latency growth based on CPU load. We fit the data to a polynomial function using the least squares method [20], which computes the *nth*-order polynomial coefficients that best fit the input data.[1] The final equations are in the legend of Figure 1.

If we compare the coefficients of the leading terms, we can see the data's growth speed. The second formula ascends 14× faster than the first formula. But the third formula escalates approximately 9× faster than the second and 125× faster than the first one – two orders of magnitude. On the range 93–100 %, if we consider a linear fit, latency grows 136×, 1207×, and 17059× faster than on previous intervals, respectively. To avoid such a rapid latency degradation on high CPU utilization, applications scale up to run on more machines after reaching a certain threshold.

---

[1] We tried multiple values of *n*s but most intervals have a good fit into linear equations (*n*=1).

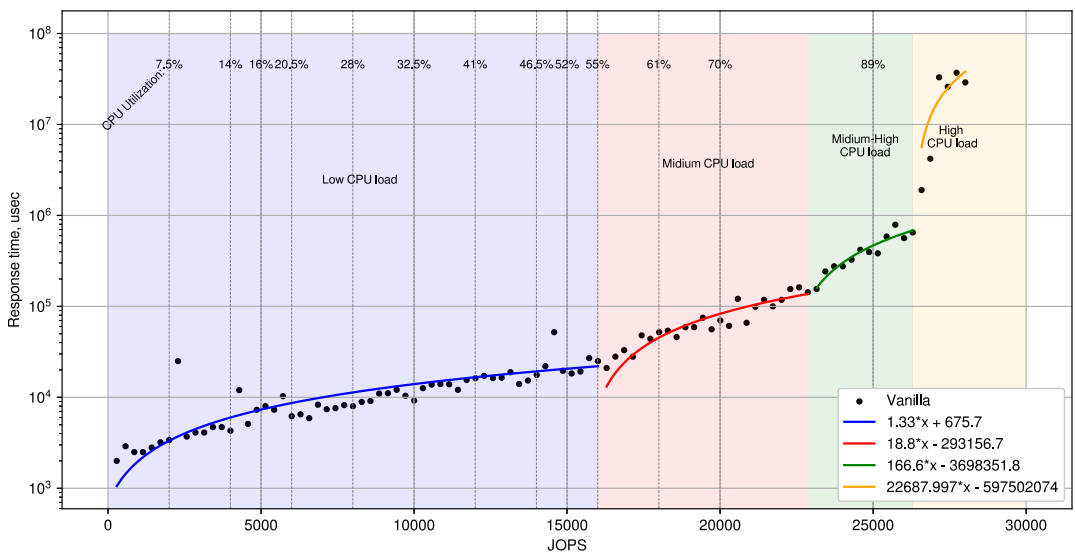

**Figure 1** An example of a response time curve for maximum latency for SPECjbb2015 together with approximate equations that describe the data on four CPU load intervals. Note the log scale. Based on latency response, we divided CPU loads into four categories: low, medium, medium-high, and high.

In a managed language like Java, GC threads (workers) co-exist with application threads (mutators) and share the same resources, especially when GC activities run parallel with the program. In OpenJDK, threads are scheduled by the underlying operating system. Workers and mutators have the same thread priority. Thus, during a concurrent GC cycle, GC workers may preempt mutator threads if there is contention on CPU resources in the system. Preempting mutators engaged in essential work, such as processing requests, leads to increased latency. This contention increases with increased CPU load, leading to early scaling out due to unsatisfied SLAs. However, workloads often do not run consistently at a high CPU load, but frequently go through periods of lower activity, when the incoming request rate is lower. If GC can be delayed to these periods, it would be possible to withstand a higher average CPU load without negative impacts on latency. This could increase the threshold for horizontal scaling, thereby improving hardware utilization.

Naturally, delaying or slowing down stop-the-world GCs might not be a good idea as stop-the-world pauses are on the critical path to the application's performance [1]. However, OpenJDK has (almost) fully-concurrent collectors, ZGC [21] and Shenandoah [12], that run concurrently with an application without pausing mutators. These collectors are not on a mutator's critical path as long as they can collect garbage at the same rate the application allocates. This has been utilized before, *e.g.,* to run concurrent GC on slower and more energy-efficient cores to reduce total energy consumption without sacrificing performance and latency metrics [27].

To this end, we propose implementing workers' scheduling exclusively during application idle intervals, aka. slack-based scheduling (slack-based scheduling later in the text) in a fully concurrent collector. We call this solution Monk as GC threads are not in a rush and run on idle cores without interfering with the rest of the world.

We hypothesize that *this technique should reduce GC pressure on CPU resources needed for mutators, which will improve latency on medium CPU loads.* This hypothesis will be validated through controlled experiments that measure latency under varying loads, as described in Section 4. Improving latency on medium CPU loads may allow delaying horizontal scaling, which is a technique often used to keep latency low as workload increases.

While the concept of slack-based scheduling is not novel and has been extensively explored in the context of real-time systems to ameliorate latency [5, 14], our study explores how slack-based scheduling of GC threads can be used to achieve better hardware utilization for server workloads.

We expect that Monk will exhibit varied behaviors under conditions of low, moderate, and high CPU loads. Low CPU load is defined by the absence of contention for CPU resources between mutators and workers. In contrast, moderate and high CPU loads are characterized by escalating levels of contention, with moderate and high loads experiencing increasing levels of contention. When the CPU load is low, we do not expect Monk to impact applications' latency since there is enough available CPU resources for workers to run on idle cores. Monk should positively affect latency on middle-range CPU utilization as contention starts growing, as long as workloads have enough periods of low CPU utilization for GC to keep the application from running out of memory. In extended periods of high CPU utilization, we expect Monk to experience GC starvation as there are fewer periods of low CPU utilization, causing GC to be unable to keep up with an application's allocation rate. To that end, we explore different fall-back policies to counter the negative effects of Monk on high CPU loads. The triggering point of a fall-back policy is a delicate choice. Falling back too late risks starving the GC, but falling back too early might lose the latency benefits from scheduling GC on idle cores. An effective fall-back policy must have optimal timing, ensuring that the latency benefits are maintained while preventing GC starvation and subsequent memory resource depletion for the application.

In this study, we investigate the following research questions:

- **RQ1** What are the effects on performance when GC is permitted to run only when there is at least one idle core to not interfere with program threads?
- **RQ2** When GC is only permitted to run when there is at least one idle core, how should we handle priority inversion, *i.e.,* when GC is on the critical path to performance?

The rest of the paper is organized in the following manner: we cover most related work and other related concepts in Section 2; Section 3 presents the Monk design and implementation; Section 4 describes our methodology; Section 5 shows our results; Section 6 describes related work in more detail.

## 2 Background

This section provides an overview essential for understanding the context of this work. We begin by exploring the Linux operating system's default scheduler, which we rely on later in our design. Following this, we survey existing literature and related work in the field, highlighting key findings and insights that inform our approach. Finally, we introduce the fundamentals of ZGC, an almost fully concurrent collector that we used to demonstrate our approach.

## 2.1 Linux OS Scheduler

The Linux OS scheduler typically refers to the process scheduler used in the Linux kernel, which is a crucial component responsible for managing and allocating CPU time to processes and threads running on a Linux-based operating system. The Linux scheduler utilizes a time-sharing, multitasking scheduling policy designed to allocate CPU time among processes.

The default scheduler of kernels from version 2.6.23 to 6.6 is the *Completely Fair Scheduler* (CFS). CFS strives to maintain a fair distribution of CPU time among tasks by assigning each task a dynamic "virtual runtime" based on its execution history, scheduling policy, and priority. The scheduler uses a combination of scheduling policies and priorities to determine which process should run next. We abbreviate this combination as simply "priority" in subsequent text, for the sake of brevity. Higher-priority tasks get preference over lower-priority ones. Each process is allocated a time quantum (or time slice) during which it can execute. Once the time quantum expires or the process voluntarily yields the CPU, the scheduler selects the next process to run.

We explored two scheduling policies: SCHED_IDLE, and SCHED_OTHER. There is an additional parameter called priority, which helps distinguishing which threads should run first of processes with the same scheduling policy. However, SCHED_IDLE and SCHED_OTHER can only have priority equal to zero. The SCHED_IDLE policy asks to be scheduled on idle cores only. SCHED_OTHER is the default Linux time-sharing scheduling.

Because the OS scheduler lacks insight into the activities and dependencies of threads within applications, it operates under the assumption that they are unrelated, except for the policies mentioned earlier. However, applications or VMs that spawn threads know if a thread is a helper thread, a GC thread, or an application thread, etc. This information could be leveraged to influence scheduling decisions.

## 2.2 Opportunistic Scheduling

Opportunistic scheduling often called slack-based scheduling has been widely explored in the context of real-time systems [5, 13, 14] but also in the context of Web programs [10]. The main idea is to schedule helper threads during applications' idle periods to achieve low latency. We discuss these works in more detail in Section 6.

### 2.2.1 Priority Inversion

Opportunistic scheduling suffers from priority inversion. Priority inversion happens in real-time and multi-priority scheduling systems when a lower-priority task temporarily holds a resource needed by a higher-priority task. Priority inversion can be a significant problem in real-time systems, for instance, where the timely execution of high-priority tasks is critical.

Roger Henriksson [14] introduces priority inheritance protocols, basic inheritance protocol [24], priority ceiling protocol [24] and the immediate inheritance protocol [19], to avoid blocking caused by priority inversion. In essence, the basic inheritance protocol dictates that when a process is waiting to acquire a lock already held by a lower-priority process, the blocking process inherits the priority of the blocked (waiting) process. The priority ceiling protocol assigns each resource a priority ceiling, representing the highest-priority process that can access it. A process can only lock a resource if its priority exceeds the ceilings. This protocol prevents deadlocks and ensures a process can only be delayed once by a lower-priority one. The immediate inheritance protocol shares requirements with the priority ceiling protocol but is simpler to implement while maintaining similar worst-case performance. Like the priority ceiling protocol, it necessitates assigning priority ceilings to resources. When a process attempts to lock a resource, its priority is immediately set to the maximum of its current priority and the resource's ceiling. This protocol also guarantees deadlock prevention.

According to Joshua Auerbach et al. [5], one approach to tackle priority inversion involves leveraging the kernel's manual priority adjustment mechanism. This entails temporarily elevating the priorities of threads responsible for delaying progress and promptly returning them to their lower priority levels afterward.

#### 2.2.2 Starvation

Opportunistic scheduling also suffers from starvation. Starvation happens when a thread with low priority is blocked from ever getting to execute. In the context of GC, if a GC worker never gets to execute, an application will sooner or later run out of memory and perform poorly or crash.

One way to address thread starvation is periodic scheduling [13] at regular intervals, allowing GC threads to preempt mutators. Another approach is to share GC work [5] between GC threads and mutators. When GC does not manage to keep up with the applications' allocation rate, mutators can switch to doing GC instead.

### 2.3 ZGC

ZGC [21] is a low-latency, parallel, concurrent, compacting GC. It implements algorithms whose stop-the-world pauses (STW) pause times do not increase with the size of the heap, including concurrent evacuation of pages during reclamation (meaning regions of memory are freed by moving all live objects away from the page). Its high-level non-generational algorithm was described by Yang and Wrigstad [35]. The GC handbook [16] briefly discusses generational ZGC.

To enable concurrent compaction, ZGC uses load barriers to trap accesses to relocated objects and remap dangling pointers to point to their updated location before accesses may commence. The ZGC algorithm forces all pointers on the heap to be remapped once per GC cycle and uses atomic operations such as compare and set to coordinate GC worker threads and program threads operating concurrently on the same objects.

As long as GC workers can keep up with a program's allocation rate, GC activities in ZGC never cause mutators to block (modulo brief STW pauses where no real GC work takes place). However, if GC falls behind, mutators will eventually stall, waiting for memory to allocate new objects.

ZGC – like all GCs in OpenJDK – runs GC threads with the same priority as program threads. For example, this would be SCHED_OTHER, the default scheduling policy on Linux. This means that GC threads can preempt mutators in CPU resource contention. This may disturb program work as demonstrated by a simplified visualization in Figure 2.

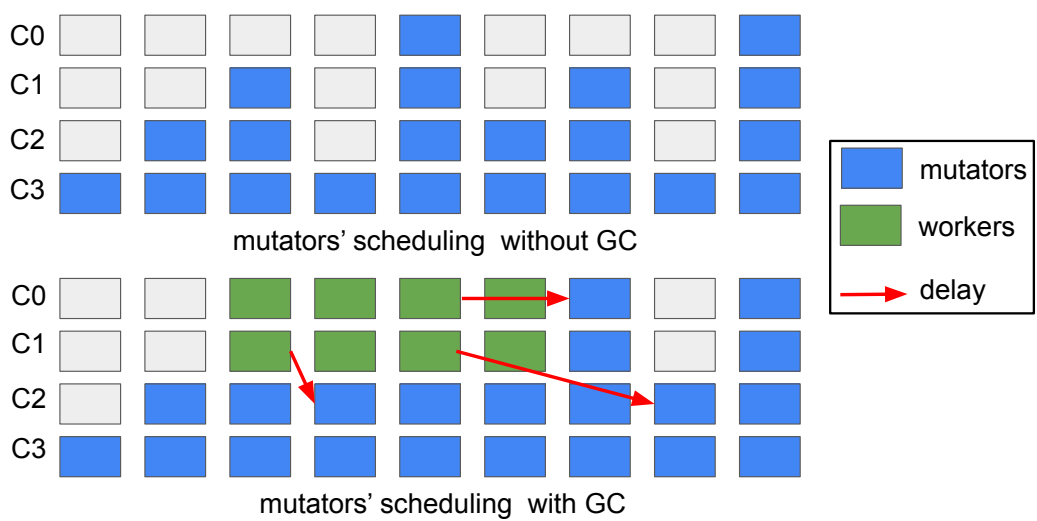

**Figure 2** Visual representation of how GC threads can delay mutators work. The upper diagram shows varying CPU loads in time due to varying incoming requests on a system consisting of four cores (C0–C3). Each column represents one point in time, and time progresses from left to right. The bottom diagram shows how mutators are affected if 2 workers with the same priority as mutators start running. Since workers can "steal" cores from mutators, some mutator tasks will be scheduled later, which causes latency delays in request processing.

#### 2.3.1 ZGC Heuristics

The goal of a concurrent collector is for the reclamation rate to match the allocation rate of the application while minimizing the impact on latency. ZGC uses non-trivial heuristics to determine when to start a GC cycle to prevent Out-Of-Memory (OOM) errors. It looks at the history of application allocation rate, current memory pressure, etc., and dynamically selects how many GC worker threads to use for every cycle, etc.

The number of GC workers is a critical parameter. To determine the appropriate number of GC workers needed to prevent OOM, ZGC analyzes the length of previous GC cycles, makes a conservative prediction of the allocation rate, and then adjusts the worker count. It then predicts the duration of the next GC cycle and calculates when the application will run out of memory at the conservatively predicted allocation rate.

If there is more time before OOM than GC duration, GC does not start. The goal is to have as few GC threads as possible not to disturb the mutator threads. However, if ZGC requests the maximum number of workers for a current cycle, it indicates that GC is critical and we are running out of time to perform GC before OOM.

### 2.3.2 Safepoints and Locks

As a concurrent collector, ZGC needs to synchronize with mutators to avoid data races on the same data. To that end, ZGC uses both locks and safepoints.

A point in a mutator's execution where it can be safely stopped by GC is called a *safepoint*. Safepoints are placed strategically in the program to minimize the impact on performance while still ensuring the correctness of garbage collection. They are typically inserted by the compiler at specific locations in the code, such as at every function/method exit, and at repeated intervals within loops. During runtime, when the JVM needs to execute a specific synchronization task (such as certain phases of garbage collection), it can instruct all relevant application threads to pause temporarily at the next safepoint. The fundamental operation of a safepoint involves the STW technique, which means that all threads executing the application are halted while the JVM performs the required background tasks.

There are several locks in the JVM that the GC grabs occasionally to coordinate with mutator threads or to coordinate concurrent GC workers. When the GC grabs a lock that is shared with mutators, the GC is on the critical path to mutator performance.

Both safepoints and locks cause priority inversion discussed in Section 2.2.1.

## 3 Design and Implementation

This section discusses the design and implementation of Monk. The implementation is based on ZGC and is written in C++.

In short, to investigate opportunistic scheduling on server workloads handling requests, we configured GC threads with the lowest priority, SCHED_IDLE, by default. We refer to this simple solution as Monk.

As briefly discussed in the introduction, the effect of Monk can be divided into three categories based on CPU utilization: low, middle, and high. When the application's CPU utilization allows mutators and GC threads to share resources without contention (low CPU utilization), there should be no differences between the default implementation of ZGC and Monk.

Monk should shine on middle-range CPU utilization as GC has enough flexibility to be postponed without starvation. Middle-range CPU utilization is crucial, as real application deployment typically run in that range to balance cost and latency. Running on low CPU utilization wastes resources, and running on high CPU utilization affects latency negatively, making it harder to guarantee meeting SLAs. Monk should allow applications to run at higher average CPU utilization by decreasing latency in the middle range before the need to scale up to meet SLAs arises.

When an application runs at a machine's full capacity, there are not enough idle resources to run GC, meaning Monk will starve GC threads. Therefore, we expect latency to increase and allocation stalls to occur more frequently in Monk as GC threads can not free garbage objects quickly. As running on a saturated machine is not a realistic deployment when trying to meet low-latency SLAs, this situation should not happen in practice on a well-configured system. However, to counteract the negative effects on a high CPU utilization spectrum when GC can not keep up with the application allocation rate, we investigated various fallback approaches (dubbed High Monk), discussed at the end of this chapter.

**Dealing with Priority Inversion** Our design should be able to handle priority inversion issues, discussed in Section 2.2.1. We increase threads' priorities if and when GC becomes critical for the application's performance/mutator progress (when GC work blocks mutators). We previously identified three potential situations:

1. GC holds a lock to data that may also be acquired by mutators.
2. GC is in a STW pause (a safepoint).
3. GC is starving, meaning it cannot keep up with the application allocation rate, eventually leading to mutators going into allocation stall.

The last situation results from machine saturation, whereas the first two occur frequently under normal operation. We consider GC to be "critical" in all three cases above and allow GC to compete with mutators for CPU resources. This translates to changing the priority of GC threads to be the same as mutators, SCHED_OTHER. So, we opted for the manual priority adjustment mechanism discussed in Section 2.2 for both priority inversion and starvation, treating all causes of GC "criticality" uniformly. In spirit, this is similar to work by Akram et al. [1] that switches GC to run at a "higher priority" (and on performance cores) during STW, while the rest of the time it runs at a "lower priority" (and on energy-efficient cores). As we implement Monk in the context of a concurrent collector without long STW pauses, we do not lower the priority of mutators.

The key challenge of implementing support for detecting and responding to GC criticality is dealing with multiple concurrent causes of priority switching. For example, multiple locks could overlap in time in a nested or non-nested fashion.

Our implementation of Monk with priority inversion uses a counter for each GC thread (Line 1 of Listing 1), initialized to zero on creation. We increment the counter when a GC thread grabs a lock or enters a safepoint and decrease it when the lock is released, or the safepoint is exited. When a thread's counter is non-zero, we increase its priority. Otherwise not. This means that different GC threads can run with different priorities as we consider their criticality individually.

The *incrementNestedCount* function (Line 8 of Listing 1) read the id of the current GC thread[2] and increments the corresponding counter. It also checks if the initial count is zero. We change the thread's scheduling policy on a change from zero to one

---

[2] The hash function can be changed to something more elaborate. Since in our experiments we have 16 GC threads and MAX_THREADS=100 a simple hash function gives a unique id.

■ **Listing 1** Common functionality for incrementing and decrementing thread local counter of criticality.

```
1  int WorkerThread::nestedCounts[WorkerThread::MAX_THREADS];
2
3  int WorkerThread::mapThreadIdToIndex(pthread_t threadId) {
4      size_t hash = std::hash<pthread_t>{}(threadId);
5      return hash % MAX_THREADS;
6  }
7
8  void WorkerThread::incrementNestedCount(WorkerThread *thread, int policy) {
9      pthread_t tid = thread->osthread()->thread_id();
10     if (WorkerThread::isZeroNestedCount(tid, thread)) {
11         WorkerThread::updateSchedulingPolicy(policy, tid);
12     }
13     int index = mapThreadIdToIndex(tid);
14     Atomic::add(&nestedCounts[index], 1);
15 }
16
17 void WorkerThread::decrementNestedCount(WorkerThread *thread, int policy) {
18     // Similar to incrementNestedCount, except priority is decreased when counter hits 0
19     …
20 }
21
22 bool WorkerThread::isZeroNestedCount(pthread_t tid, WorkerThread* thread) {
23     int index = mapThreadIdToIndex(tid);
24     return Atomic::load(&nestedCounts[index]) == 0;
25 }
```

■ **Listing 2** A simplified example of the updated ZLock lock function.

```
1  inline void ZLock::lock() {
2      if (Thread::current()->is_gc()) { // True if current thread is a GC worker
3          WorkerThread::incrementNestedLockCount(Thread::current(), SCHED_OTHER);
4      }
5      _lock.lock();
6  }
```

(SCHED_OTHER) or from one to zero (SCHED_IDLE) through sched_setscheduler Linux function.

When GC is starving, the thread that coordinates GC (the director thread) iterates over the counters in the array and increments them by one. Thus, there is possible contention on a thread's local counter, for example, if a worker thread grabs a lock simultaneously. Thus, increments and decrements on the counters must be atomic. Atomic instructions help with data races but cannot guarantee a specific order. For example, one thread observes that GC stops being critical and wants to decrease its priority. At the same time, the other thread observes that a GC thread becomes critical and wants to increase the GC priority. Even though the increment and decrement of a counter are serialized, we can not guarantee that changing threads' priority will

**Listing 3** A simplified examples of updated safepoint function.

```cpp
class IncrementLockClosure: public ThreadClosure {
  virtual void do_thread(Thread *thread). {
    WorkerThread::incrementNestedLockCount(thread, SCHED_OTHER);
  }
};

void ZCollectedHeap::safepoint_synchronize_begin() {
  IncrementLockClosure increment_one;
  _heap.gc_threads_do(&increment_one); // Apply increment_one to all GC threads
  ...
}
```

happen in the right order. Rather than wrapping the counter and priority operations (which involves a costly system call) in a lock, we use a helper thread that periodically checks that thread's priority aligns with their counter value. It does not guarantee that GC threads always run with the right priority, but running with the wrong GC priority will not affect the correctness of an application, only performance parameters. We run such a check every 10 ms, which, in a 2.5 h run of SPECjbb2015, gives us 900 000 opportunities to correct mistakes, and we rarely see more than 45. Of course, evoking a helper thread more often will reduce the time we run with a wrong GC priority, but it is also more costly. We defer the decision on whether a lock or a more frequent helper thread invocation presents a more favorable cost-benefit trade-off to future investigations.

We updated specific lock functions that ZGC uses exclusively to test if switching a scheduling policy when GC holds a lock is critical for performance. A slightly simplified code is depicted in Lines 2 to 4 of Listing 2, where two things are shown: (1) only the current GC thread updates its scheduling policy and (2) switching scheduling policy adds overhead to each lock and unlock operation, which can be a major source of unnecessary overhead if ZGC does not compete on the lock with a mutator.

The function *safepoint_synchronize_begin* in Listing 3 shows how we iterate over all GC threads (line 10) and update their scheduling policy using a helper lambda function (line 3).

We experimented with different fallback thresholds to prevent GC starvation during high CPU utilization. We call these approaches "High Monk" (Listing 4). If we are inside a critical section, we decrease GC priority and end the critical section (lines 3–5). Otherwise, we check if we should start a critical section. This means we fall back to Vanilla behavior every second of the GC cycle when GC keeps on a critical path. If we start a critical section, we read the current CPU utilization (line 7, more details below). If the CPU utilization is higher than the threshold provided as a command line parameter (HIGH_CPU) and if ZGC wants to use the maximum number of allowed GC threads (which indicates GC pressure), we increase the priority of all GC threads (lines 9–11).

We also experimented with the implementation where we kept Vanilla's behavior for as long as GC was on the critical path (see Appendix A). We will not present this

■ **Listing 4** High Monk

```
1  // we come here if zDirector has decided to start a GC cycle
2  if (ZHeap::heap()->is_inside_critical_section()) {
3    DecrementLockClosure my_closure_decrease;
4    ZHeap::heap()->gc_threads_do(&my_closure_decrease); // Decrease priority for all threads
5    ZHeap::heap()->set_inside_critical_section(false);
6  } else {
7    unsigned long long cpu_usage = get_current_CPU_utilization(); // read /proc/stat
8    if (cpu_usage > HIGH_CPU && number_of_workers == MAX) {
9      IncrementLockClosure my_closure_increase;
10     ZHeap::heap()->gc_threads_do(&my_closure_increase); // Increase priority for all threads
11     ZHeap::heap()->set_inside_critical_section(true);
12   }
13 }
```

solution in detail, as the alternative approach of executing every other cycle demonstrates better latency with minimum throughput hit. However, we want to emphasize that this is another point in the design space that requires further exploration. We do not claim that the every-other-cycle method is the optimal solution; instead, we highlight that experimenting with this parameter might be critical for production solutions.

As mentioned above, we control different fallback thresholds with a HIGH_CPU flag set at the command line. This value denotes when GC becomes critical, and Monk needs to fall back to the SCHED_OTHER GC thread scheduling. Consider a scenario where a machine has 10 cores, all of which are utilized. In this context, when we set HIGH_CPU=100 %, it means that High Monk will operate similarly to Monk up until no cores are available for GC before the commencement of a GC cycle. With HIGH_CPU=90 %, High Monk reverts to default behavior (*i.e.,* GC workers run with normal priority) when only one core is available before the start of a GC cycle. We read the current CPU utilization from the /proc/stat file, which we read each time a decision about a GC cycle is made (not shown in Listing 4 for simplicity). From the /proc/stat file we get four following values:

- cpu_user – time spent in user mode,
- cpu_nice – time spent in user mode with low priority (nice),
- cpu_system – time spent in kernel/system mode,
- cpu_idle – time spent idle.

To calculate current CPU utilization, we use the following formula. First, we calculate current CPU usage:

$$cpu\_sum = cpu\_user + cpu\_nice + cpu\_system + cpu\_idle$$

Next, we calculate the delta between two reads:

$$cpu\_delta = cpu\_sum - cpu\_last\_sum$$

Then, we compute the idle time delta and time spent working:

$$idle\_delta = cpu\_idle - cpu\_last\_idle \qquad cpu\_used = cpu\_delta - idle\_delta$$

And finally, we calculate current CPU usage in percent:

$$\text{cpu\_usage} = 100 \cdot \text{cpu\_used}/\text{cpu\_delta}$$

We calculate a decaying average to smooth CPU utilization values.

Table 1 lists all the tested configurations. MONK is Monk, the configuration where GC threads are always scheduled on idle cores. If a configuration switches scheduling policy inside safepoints, we write its name with a _S suffix, like MONK_S and HMONK_S_C{0–4}. If it switches a priority inside locks, it has _L suffix, like MONK_L. High Monk has a "H" prefix and a _C*N*, suffix, where *N* is the number of available cores before a GC cycle when High Monk backs off and uses SCHED_OTHER scheduling priority. Examples are **HMONK_S_C{0–4}** and **HMONK_C{0–4}**

**Table 1** Explanation of Configurations

| Configuration | Description |
|---:|---|
| MONK | Monk configuration with GC threads on idle cores |
| MONK_S | Monk with scheduling policy change in safepoints |
| MONK_L | Monk with scheduling policy inside locks |
| HMONK_S_C{0–4} | Monk with scheduling policy change in safepoints and when {0,1,2,3,4} available cores |
| HMONK_C{0–4} | Monk with priority change when {0,1,2,3,4} available cores |

## 4 Methodology

In this section we describe our methodology.

### 4.1 Hardware and Software

We evaluate our work by comparing our modified ZGC with its unmodified baseline, which we refer to as "vanilla". The baseline is generational ZGC in OpenJDK 21.

We tested two different architectures from two vendors to see if the results were transferable. First, Intel Xeon *SandyBridge* EN/EP server machine running Oracle Linux Server 8.4 with kernel 5.4.17. The machine has 32 identical CPUs with 2 hardware threads per core, which we configured as a single NUMA node to avoid NUMA effects. The CPU model is Intel Xeon CPU E5-2680 running at a maximum of 2.7 GHz with 64 KB L1 cache, 256 KB L2 cache, a shared 20 MB L3 cache, and 30 GB RAM.

Second, an AMD machine running Ubuntu 22.04 with kernel 6.5.0-27-generic. The machine has 32 identical CPU cores with 2 hardware threads per core and a single NUMA node. The CPU model is Ryzen Threadripper 3970X running at a maximum of 3.7 GHz with 64 MB L1 cache, 512 MB L2 cache, 1024 MB L3 cache, and 256 GB RAM.

We used cgexec command to run benchmarks inside one control group to ensure that background tasks do not interfere with GC work since GC work has the lowest priority. We configured the group to run on 30 cores to reserve 2 cores for other background tasks.

### 4.2 Benchmark: SPECjbb2015

To obtain a comprehensive understanding of MONK we used SPECjbb2015 [30]. As we are interested in delivering low latency, we are more interested in concurrent rather than parallel workloads. Parallel workloads span many threads simultaneously to saturate the machine as much as possible, which does not leave space for GC threads. Due to their asynchronous nature, concurrent workloads tend to vary more, for example, due to different requests in different time windows. Another useful characteristic of SPECjbb2015 is that it tests the machine at different CPU loads, allowing us to see our techniques' behavior on all three CPU load ranges.

SPECjbb2015 is a long-running benchmark, which simulates an online store under increased pressure for about 2,5 hours. The benchmark gradually increases the number of incoming requests until the machine is fully saturated. The benchmark calculates two aggregate metrics: critical-JOPS and max-JOPS. JOPS stands for Java operations per second. Critical-JOPS represents latency scores as a geometric mean of multiple values of response times (10 ms, 25 ms, 50 ms, 75 ms, and 100 ms). Typically, the 10 ms and 25 ms response times are the most relevant for a low-latency GC such as ZGC. Critical-JOPS for 10 ms will show the maximum number of simultaneous requests (JOPS) the system handled under 10 ms.

Max-JOPS is a throughput metric reflecting the volume the machine can handle under maximum load.

To connect these performance metrics to Figure 1, draw the imaginary horizontal line at 100 ms ($10^5$ in log scale), and when reaching the first black dot that is above the line, the corresponding JOPS (a bit more than 20 000) indicate how many requests were handled under 100 ms latency. 100 ms latency is the highest latency value reported by SPECjbb2015. max-JOPS in this example is around 27 000 JOPS.

The benchmark consists of three phases: warm-up, response curve building, and profiling.

During the warm-up phase, the benchmark actively engages the system to determine its preliminary maximum capacity in handling requests. This phase is characterized by high CPU load as the benchmark stresses the system to identify its performance limits.

Following the warm-up phase, the benchmark transitions to building the response curve. During this phase, the benchmark divides the range from zero to the preliminary maximum into steps, gradually increasing the number of requests and measuring their response times. Concurrently, latency-related performance metrics such as critical-JOPS are measured. During this phase, CPU usage gradually escalates from low to high as the workload intensifies.

As the machine reaches maximum load, the max-JOPS score is calculated.

Figure 3 provides visual representations of the dynamic worker count (Figure 3a) and CPU usage (Figure 3b) for SPECjbb2015. Figure 3a shows that ZGC initially allocates a high number of workers during the warm-up phase, and then gradually increases it from one worker as the workload grows. Meanwhile, Figure 3b illustrates the CPU usage of SPECjbb2015 before GC. Note that CPU usage does not reach 100 % due to the utilization of 30 out of 32 available cores (approximately 94 %). CPU usage peaks intermittently during the warm-up phase. Towards the end of the benchmark, sustained periods of full CPU capacity are observed, indicating a higher risk of GC starvation if GCs are only allowed to be scheduled on idle cores.

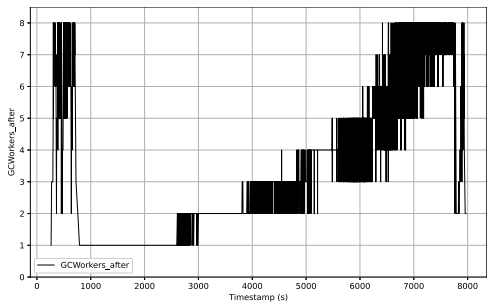
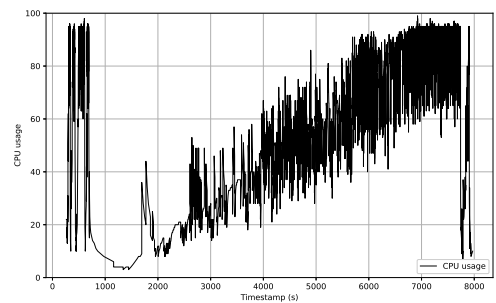

**(a)** The dynamic number of GC workers for SPECjbb2015

**(b)** CPU usage of SPECjbb2015 taken from /proc/stat before each GC cycle

■ **Figure 3** Dynamic GC Workers and Corresponding CPU Usage in SPECjbb2015

## 4.3 Benchmark: Hazelcast

Hazelcast is a framework designed for real-time stream processing. Hazelcast's workload and, consequently, CPU load are determined by the size of its key set. CPU load increases with the increments in the key-set sizeFigure 4b, similar to how CPU load increases with increased JOPS for SPECjbb2015. However, it is important to note that the CPU load in Hazelcast fluctuates more than in SPECjbb2015. This observation will be important when we explain the results. We conducted experiments to explore the impact of different workloads using key-set sizes from 80 000 to 500 000. Figure 4a shows the dynamic number of GC workers for a few key-set sizes. As observed, the number of GC workers and frequency of GC directly correlate with CPU utilization with higher key-set sizes using more GC workers and CPU resources.

For the rest of the parameters, we followed suggestions by Topolnik [29].

## 4.4 Benchmarking Methodology

We run SPECjbb2015 using 5 JVM instances, which lets us identify performance anomalies and outliers that might not have been discernible using a single JVM instance. Since SPECjbb2015 is a long-running benchmark(2.5 h), we expect the time of Just-In-Time and C2 compilation to be notably lower than the benchmark's execution time, which allows the benchmark to reach a steady state for performance measurements. Reaching a steady state is crucial for Java benchmarking since it has high jitter in

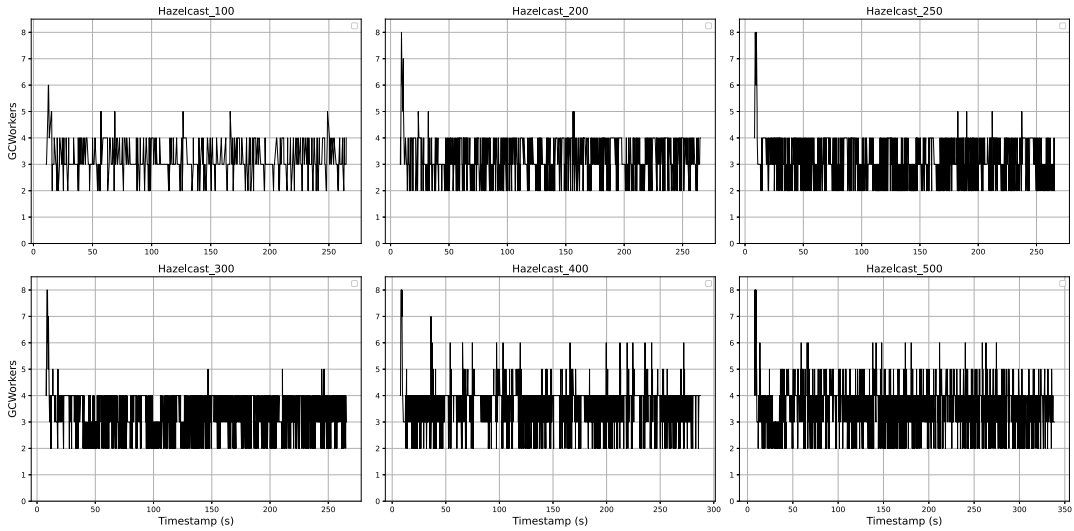

**(a)** The dynamic number of GC workers for Hazelcast and different key-set sizes.

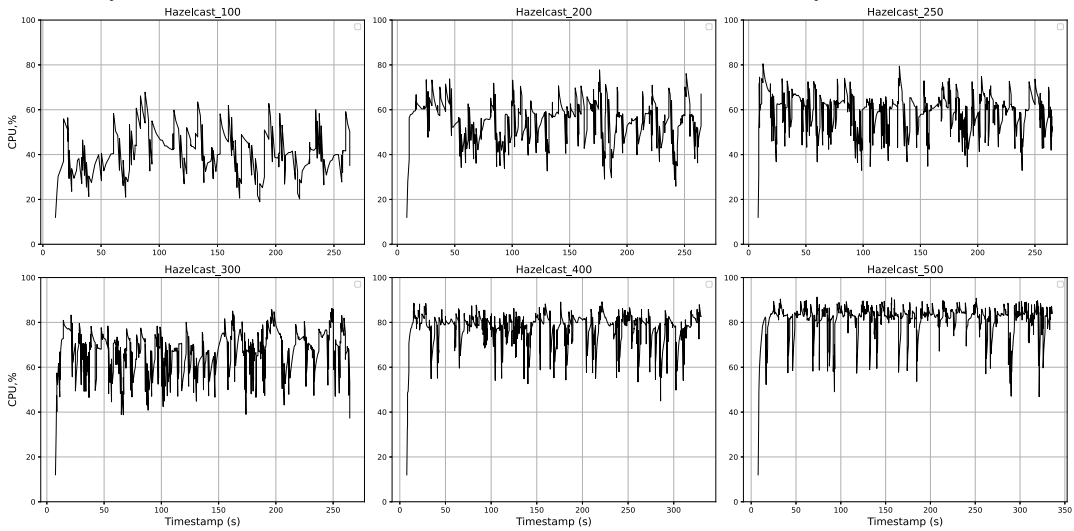

**(b)** CPU usage of Hazelcast taken from /proc/stat before each GC cycle

**Figure 4** Dynamic number of GC Workers and Corresponding CPU Usage in Hazelcast

the JVM start-up/system warm-up phase, so multiple sources [9, 18] recommend accounting for it.

Regarding memory, we ran SPECjbb2015 with 26 GB of memory on both Intel and AMD. We opted to run with the maximum possible heap size since SPECjbb2015 benefits from all available resources, allowing the benchmark to reach higher performance scores. However, we must also store data and ensure that other tasks have enough memory to run OS smoothly. We used 26 GB on the Intel machine since the memory there was limited. We used the same amount of memory on AMD so that memory was not different between the experiments on the two machines.

We run Hazelcast 30 times, as latency variation is higher than in SPECjbb2015. The benchmark runs several minutes, which is notably longer than JIT and C2 phases, so the benchmark should be able to reach a steady state. We use 8 GB of memory.

We used different statistical tests to verify the results and draw conclusions. We employed Welch's t-test [33] to determine whether the differences between the means of the compared results from vanilla and MONK are statistically significant. Welch's t-test is particularly useful in cases where we can not make assumptions about the shape of data distribution and the variances of the compared groups are unequal. We used the Shapiro–Wilk test [25] to check if the data set has a normal distribution.

We used the Median Absolute Deviation (MAD) [15] method to identify and remove any statistical outliers from our data set without normal distribution. Outliers can significantly skew the results of statistical tests, leading to misleading conclusions.

## 5 Results

In this section, we present the findings and outcomes of our experiments. We provide an analysis of the data collected and discuss the implications of our results.

### 5.1 Monk

This section answers **RQ1**: *What are the performance effects of permitting GC to run only when there are idle cores?*

Figure 5 compares one run of MONK implementation to unmodified ZGC (vanilla) for both SPECjbb2015 and Hazelcast. Note the log scale. This graph serves an explanatory purpose. For statistical comparison, please refer to Table 2.

Let us look at the upper graph, which is for SPECjbb2015. As predicted, both MONK and vanilla have a similar response time on low CPU utilization in the range of 0–59 % (vertical lines), as in both cases, GC and mutators can share the system without contention.

On middle CPU utilization in the range 60–80, two curves start departing in response time, with MONK showing better response time (orange dots are consistently lower than black). On middle-range loads, contention between mutators and GC for CPU resources increases, and scheduling GC on idle cores releases CPU pressure for mutators.

**Monk: Opportunistic Scheduling to Delay Horizontal Scaling**

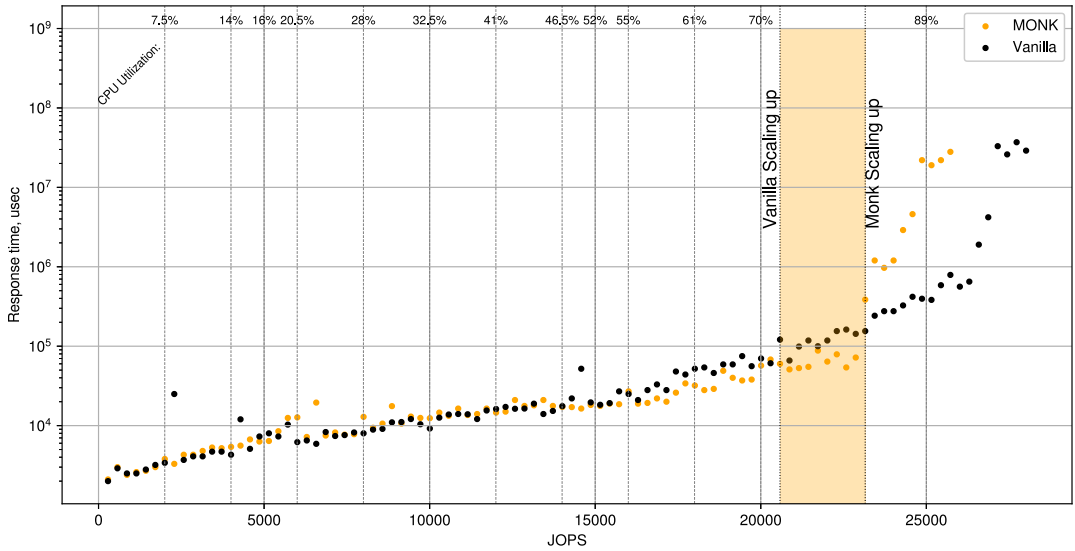

■ **Figure 5** Maximum response time in log scale at different JOPS (injection rate/number of requests per second) for *a single* run of vanilla and Monk for SPECjbb2015 benchmark. Orange color showcase Monk. Black represents vanilla. Due to GC starvation, Monk can not execute above 90 % CPU utilization. The dotted black vertical lines show a scaling-up moment for unmodified ZGC and Monk if the maximum latency SLA is below 100 ms. Vertical lines in plots **a** show CPU utilization. CPU utilization grows together with increased JOPS.

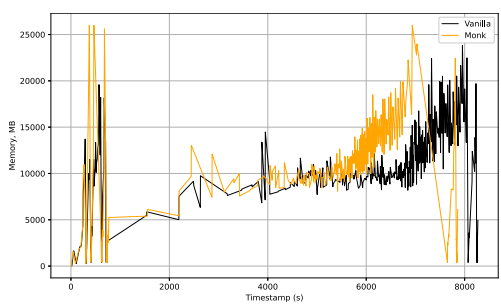

**(a)** Memory after GC in MB during SPECjbb2015's execution for both vanilla and Monk configurations. The maximum capacity is 26 GB.

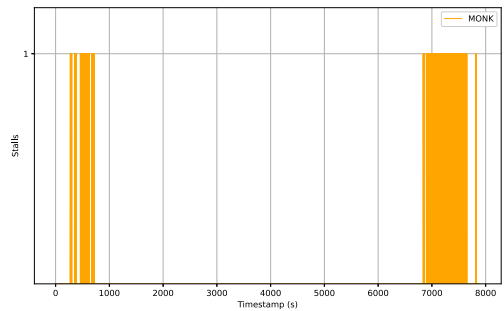

**(b)** Stalls during SPECjbb2015's execution for Monk. The graph represents each stall event with a value of 1, while the absence of stalls is represented by 0.

■ **Figure 6** Memory and stalls of unmodified ZGC (vanilla) and Monk configurations for SPECjbb2015 benchmark. Orange color showcase Monk. Black represents vanilla.

At high CPU utilization in the range of 80–100, MONK starves GC, meaning GC does not get enough CPU resources to keep up with the allocation rate, so the response time for MONK quickly degrades (orange dots are noticeably higher than black).

As expected, MONK's response time during periods of high CPU utilization is worse than vanilla's. As expected, MONK has allocation stalls – an average of 289 693 on high CPU loads (Figure 6b). However, up until some point, MONK's functionality can effectively delay the need for horizontal scaling measures. Assume that scaling would

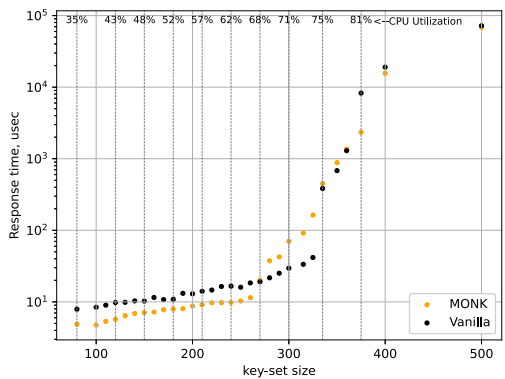 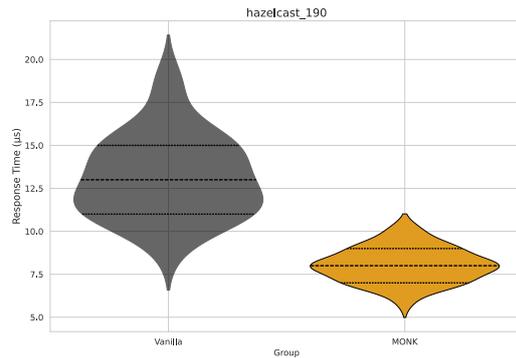

**(a)** Maximum response time at different key-set sizes for aggregated 30 runs of vanilla and Monk for Hazelcast benchmark. Vertical lines show CPU utilization. CPU utilization grows together with increased key-set size.

**(b)** Violin plots of response times for vanilla and Monk configurations for Hazelcast with the key-set size 190 000.

**Figure 7** Latency comparison of unmodified ZGC (vanilla) and Monk configurations for Hazelcast benchmark. Orange color showcase Monk. Black represents vanilla.

happen when maximum latency is above 100 ms. In vanilla, it happens at around 20 000 JOPS and approx. 72 % CPU load (first solid black vertical line Figure 5). MONK can satisfy this SLA until approx. 80 % CPU utilization(second solid black vertical line Figure 5), so the scaling to more servers can happen later(indicated by the orange shadow in Figure 5).

We also want to explicitly acknowledge the memory tradeoff associated with postponing GC. Delaying GC work may result in not immediately reclaiming memory allocated by mutators. Figure 6a illustrates that Monk consumes more memory during the benchmark's execution reaching the maximum memory capacity (26 GB) sooner than vanilla. SPECjbb2015 halts the construction of the response curve once the system reaches its maximum capacity, subsequently measuring the max-JOPS metric. Thus, the max-JOPS metric inherently reflects memory usage differences, as this metric would be higher if the application could run for a longer duration. That said, it is challenging to quantify the exact amount of extra memory required by Monk, as it depends on the nature of the benchmark, including factors such as the extent of GC delay and the available CPU resources.

Now, let us look at the aggregated results of standard performance metrics (max-JOPS and critical-JOPS) for 5 runs presented in Table 2. MONK has approx. 6 % lower throughput indicated by lower max-JOPS. This is expected as MONK quickly degrades response time on high CPU loads. However, regarding latency, represented by critcal-JOPS, especially latency below 25 ms, MONK can handle ≈13 % more requests and still satisfy the SLA.

With this said, we will explore priority inversion techniques to bring MONK's response time on high CPU loads down to the vanilla level in the next section.

But before we move on, let us look at the second benchmark. Hazelcast's curve (Figure 7a) closely resembles that of SPECjbb2015 (Figure 5). However, due to the significant variation in CPU utilization for each key-set size, the benefits of Monk are evident even with lower average CPU utilization. Unlike the vanilla approach, this is because Monk avoids disturbing mutators during CPU spikes. In SPECjbb2015, the CPU load increases more smoothly, so mutators do not compete with GC for cores at lower average CPU loads. In contrast, Hazelcast's spikier nature means that mutators and GC are more likely to compete even at lower average CPU utilization. As a result, mean latency is lower for Monk on lower average CPU utilization, and latency variance is lower as shown on Figure 7b.

Referring to Table 3, we observe that the mean latency of Monk is up to 40 % lower than that of the vanilla version at low CPU utilization. However, this advantage diminishes with larger key-set sizes, similar to the behavior observed with SPECjbb2015 under higher CPU loads. Interestingly, latency appears to improve again when the benchmark saturates the machine in the hazelcast_(375–500) range. At this point, the benchmark is misconfigured, as it should not run at total machine capacity. In this extreme scenario, both versions struggle to maintain performance, making predicting which one will perform better difficult.

## 5.2 Priority Inversion

Now, we answer the **RQ2**: how to handle priority inversion. We separated this problem into two subproblems: (1) locks and safepoints and (2) GC starvation. (1) happens at any CPU utilization, while (2) is only a problem on high CPU loads.

### 5.2.1 Locks and Safepoints

An important question to answer is *"What are the performance implications of inversing priority for locks and safepoints?"*.

As discussed above, if GC is inside a stop-the-world pause, mutators can not execute, so these pauses should be processed as fast as possible to resume mutators. However, if mutators are blocked from executing, potentially more cores are idle. Additionally, switching priority has an overhead and it is unclear if the overhead is greater than the benefit.

The same is true for GC locks on shared data. We do not want GC to hold locks forever. However, there are two nuances: (1) a lock on shared data does not mean that a mutator necessarily accesses it and therefore gets blocked, (2) as long as there are enough idle cores for GC threads to execute without interruption switching priority might not be necessary.

The aggregated results in Table 2 show that lock priority inversion (MONK_L) creates a lot of overhead to the point that 10 ms latency gets worse by almost 60 %. The payoff increases on higher latencies (from 0.41 for 10 ms to 0.95 for 100 ms) but never reaches the same efficiency as MONK. It is possible that a different strategy for lock-based switching would have yielded better results. For example, waiting until a mutator blocks to change priority will avoid the overhead of the unnecessary changes mentioned above, but on the other hand, would never avoid the blocking completely. We leave this for future work.

Inverting priority inside safepoints (see Table 2, MONK_S) seems to introduce some overhead on low latencies (10 ms, 25 ms) compared to MONK ($\approx 1\%$), evens up with MONK for 50 ms latency and gets slightly better for 75 ms and 100 ms ($\approx 1\%$). This could be because low latencies are measured on low CPU loads where there are enough idle cores and investing priority introduces unnecessary overhead. Contrary, 75 ms and 100 ms are measured later on higher CPU loads, so inverting priority allows GC to complete stop-the-world pauses faster than MONK since GC can get cores with less waiting time.

Since lock priority inversion seems inefficient, in the next set of experiments we based our solution on MONK and MONK_S only.

### 5.2.2 GC Starvation

In this section, we explore how performance parameters are affected if we fallback to the default scheduling policy when high CPU loads are detected.

Depending on the machine, ZGC dynamically decides on the maximum number of GC threads. By default, which we did not change, it picks 25 % of the number of cores as a number of GC threads for young and old generations. Since in our experiments, both machines had 32 cores, ZGC sets the maximum number of GC threads to 8. Consequently, this is the number ZGC requests when the system runs out of time to recycle memory before OOM, in other words when GC is pressing. We tried to rely on this metric alone, but since ZGC estimates time conservatively, it started requesting the maximum allowed number of threads very early. This caused us to revert to default scheduling too early to remove MONK's benefits on middle-range CPU loads.

Thus, as discussed in Section 3, in addition to checking for the maximum requested workers, we measured CPU load before each GC cycle to estimate how many idle cores are left in the system. Since the maximum number of GC threads in our experiments is 8 (25 % of cores, by default) and our cores support hyperthreading, we tested multiple configurations based on the number of idle cores left in the system from 0 to 4 as a fallback threshold. It is important to understand that once we fall back to the default scheduling policy, GC can take up to 8 cores shared with mutators following CFS Linux scheduling rules. So 0–4 cores is just an indication of possible GC starvation.

As mentioned above in Section 5.2.1, we added the High MONK solution on top of MONK and MONK_S.

**High Monk with addressed Safepoint Priority Inversion**   The results in Table 2 indicate that High MONK using safepoints to avoid priority inversion sees almost no regression in max-JOPS (except HMONK_S_C0), but significantly improves multiple critical-JOPS values. For higher latencies, starting earlier is better. But for shorter latencies, the reverse is true.

On Sandy Bridge, 100 ms critical-JOPS improve from 3 % to 6 % with more cores as heads-up. 50 ms and 75 ms latency also benefits from a heads-up both improving up to 8 %. As expected, lower 25 ms latency has a reverse trend, starting falling back later is a better option, with improvements going from 4 % for HMONK_S_C4 to 11 % for HMONK_S_C0.

The results hold true on AMD as well. Overall trends are the same especially on higher latencies, although with values a couple of percent less. 25 ms latency does not seem to be affected by the falling back strategy, where all but one performs by 8 % better than vanilla. But 10 ms latency has a more clear trend towards preferring a later fallback.

So a clear trade-off is exposed here: if you want to optimize for lower latencies, falling back needs to start later, while if one optimizes for higher latencies falling back should happen earlier.

Unfortunately, the data for 10 ms latency is inconclusive due to high variation, which is common for tight latency values. We will return to this in Section 5.4.

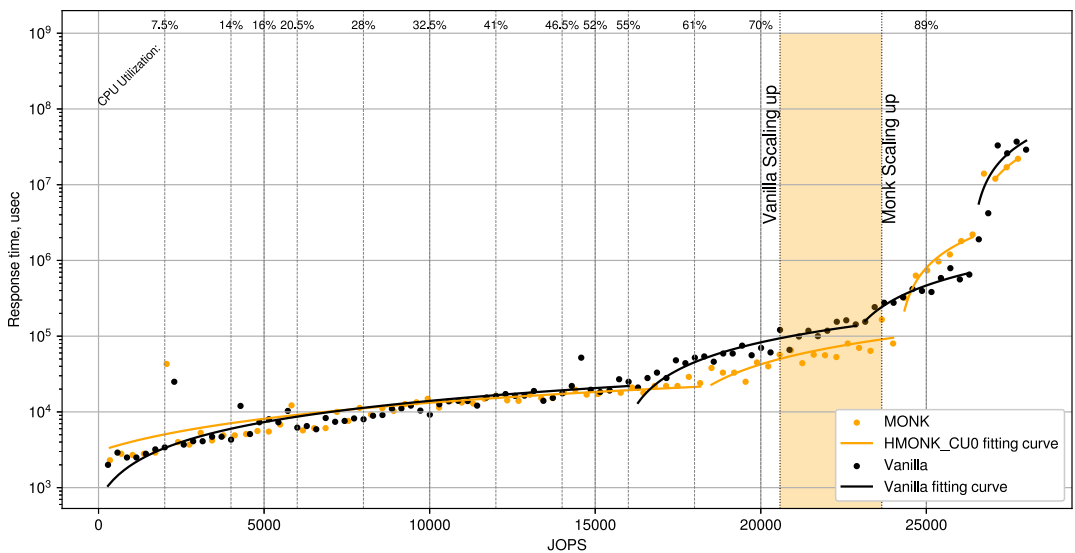

**Figure 8** Maximum response time in log scale at different JOPS (injection rate/number of requests per second) for *a single* run of unmodified ZGC (vanilla – black dots) and HMONK_C0(orange dots) for SPECjbb2015 benchmark. Vertical lines show CPU utilization. CPU utilization grows together with increased JOPS. Solid lines show how data fits into linear equations. Black lines are for vanilla and orange lines are for HMONK_C0.

**High Monk** Table 2 show the results for HMONK based on MONK. On both Intel and AMD, we have some indication that HMONK works better on short latencies (25 ms) than HMONK_S, showing the same pattern MONK and MONK_S have. For example, on average, HMONK_C0 on Sandy Bridge improves critical-JOPS on 25 ms latency by 15 % while HMONK_S_CU0 by 11 %. We see a similar trend on AMD, HMONK has 1 % latency improvement compared to HMONK_S.

Figure 8 shows how this implementation behaves compared to our baseline, unmodified ZGC. As expected, we see three distinct phases: low, middle, and high CPU. On low CPU loads, both unmodified ZGC and HMONK_C0 follow each other, although HMONK_C0 fits the same curve longer. On the middle ranges, the two curves diverge:

HMONK_CU0's equation is $y = 13.2 \cdot x - 221\,508.6$ with the leading coefficient being by 1.42× (20 %) lower than baseline. The response time of HMONK_CU0 on high CPU loads is worse than vanilla on high CPU loads, but it is considerably better than MONK in Figure 5, which could not reach max-JOPS values higher than 25155, while HMONK_CU0 reaches above 26 000.

Unfortunately, the comparison of HMONK and HMONK_S does not show statistical significance on most metrics. So, we can not conclude that one version is better.

We also tested the implementation of High Monk on Hazelcast. The results are similar, and we leave them out for brevity. Although for Hazelcast HMONK_C4 shows the same performance as Vanilla. This is expected as Hazelcast needs more headroom due to its runtime characteristics, like CPU utilization and spikiness.

### 5.3 Transferability of Results

This section discusses Monk's behavior across different systems.

As outlined in Section 4.1, we tested SPECjbb2015 on two distinct architectures from Intel and AMD with slightly different OS. Overall, our results show consistency: addressing locks' priority inversion incurs a notable expense on both machines, while tackling priority inversion stemming from safepoints and GC starvation appears more advantageous.

However, a consistent trend in which the AMD machine demonstrates slightly less improvement than Intel is worth noting. We suspect this disparity may stem from differences in cache sizes. GC gives rise to cache pollution, where loads in GC threads (typically not reused) evict loads from mutators. We suffer less from cache pollution by running GC less in MONK configurations. When the caches are larger, this effect is less pronounced. With significantly more cache memory, the AMD machine enhances the performance of our baseline, resulting in smaller differences between Monk and vanilla.

Additionally, we observed that HighMonk needs more headroom on the AMD platform than Intel. For instance, as shown in Table 2, configurations like HMONK_C2 or HMONK_S_C2 perform best on AMD for low latencies, whereas HMONK_C0 or HMONK_S_C0 excel on Intel. We attribute this variation to differences in CPU frequency, with AMD's higher frequency leading to a greater allocation rate for mutators. Consequently, the GC must fallback earlier to maintain pace with the application's allocation rate.

Despite these distinctions, we believe these variations do not undermine the transferability of Monk. Instead, they emphasize the importance of caution when applying this technique across multiple architectures.

### 5.4 Statistical Analysis

This section undertakes a statistical analysis. In particular, we will demonstrate the stability of the data in Table 2 and the results of Welch's t-test described in Section 4.4.

Table 2 shows the relative standard deviation (RSD) of the data from 5 JVM invocations (5 runs of SPECjbb2015). Overall, it does not get higher than 3.05 % on Sandy Bridge and 1.69 % on AMD. However, for 10 ms latency RSD is considerably higher than for other metrics. We suspect that this is because, at such a low latency, "flukes" become more noticeable, which transfers to the stability of the results. As a takeaway, SPECjbb2015 seems to have stable performance measurements, except for 10 ms latency. So, if 10 ms is an important value, a different benchmark should be considered.

Table 2 lists p-values for statistical significance. We use a p-value equal to 0.05 to claim the statistical significance of our data. However, we prefer to report these values as the 0.05 value is not universal, and its usage is not without criticism [22, 32]. But overall, most of the values in Table 2 are either way below(<0.00) or far above 0.05, allowing us to confirm or reject the null hypothesis more confidently.

Table 3 reports a very high RSD for Hazelcast, which sometimes reaches higher than 70 %. However, most of the p-values are below 0.00, which allows us to compare the results.

## 6 Related Work

This work explores the implication of slack-based GC scheduling for a fully concurrent ZGC. Therefore the rest of this section describes in greater detail studies related to slack-based scheduling. At the end of this section, we will emphasize the main differences between our and previous works.

Joshua Auerbach et al. [5] introduce the Metronome-TS (Tax-and-Spend), an extension of IBM's Metronome real-time JVM. The "Tax" aspect of the paper delves into the concept of requiring mutator threads to contribute to garbage collection efforts during their idle periods. This stems from the realization that relying solely on concurrency may not suffice for real-time systems, as it hinges on the underlying operating system for scheduling, which can introduce undesirable lag. The paper also investigates slack-based scheduling, where garbage collection tasks are executed during periods when a mutator voluntarily relinquishes CPU resources. Typically, these threads have lower priority, but they cannot guarantee meeting specific timing requirements. A novel approach is presented by combining "Slack" and "Tax". Background GC threads run at lower priority during CPU idle periods, collecting Minimum Mutator Utilization(MMU) and performing some garbage collection tasks. Mutators alternate between garbage collection and their regular work, stepping in only when the "Slack" GC does not sufficiently address the required tasks. This coordination is ensured through a banking system, allowing mutators to withdraw credits and assess whether they need to engage in garbage collection. Tomas Kalibera et al. [17] compare performance and schedulability of different approaches to garbage collection scheduling on uniprocessor systems, *e.g.,* slack-based [14], periodic [6, 7], and hybrid [5]. GC runs on a separate thread to take advantage of the operating system scheduler.

Thomas Gerlitz et al. [13] present the development of a reference-counting soft real-time garbage collector tailored for Android, in contrast to traditional mark-and-sweep methods that necessitate suspending mutators during heap traversal for object marking. This contribution extends Android 2.2 to imbue it with real-time capabilities,

**Table 2** Aggregated performance metrics of 5 runs of SPECjbb2015 normalized to unmodified ZGC – higher is better. Green indicates statistically better, red indicates statistically worse, and white shows the results without statistical significance. HMONK_Co* has 4 runs. p-value is the result of the Welch test. RSD is the relative standard deviation in %.

| HW | Conf | max-JOPS | | | 10 ms | | | 25 ms | | | 50 ms | | | 75 ms | | | 100 ms | | | critical-JOPS | | |
|---|---|---|---|---|---|---|---|---|---|---|---|---|---|---|---|---|---|---|---|---|---|---|
| | | val | p-val | RSD | val | p-val | RSD | val | p-val | RSD | val | p-val | RSD | val | p-val | RSD | val | p-val | RSD | val | p-val | RSD |
| Intel | vanilla | 1 | – | 0.54 | 1 | – | 2.10 | 1 | – | 1.35 | 1 | – | 1.09 | 1 | – | 1.40 | 1 | – | 0.69 | 1 | – | 1.10 |
| | MONK | 0.94 | 0.00 | 0.96 | 1.01 | 0.19 | 2.04 | 1.14 | 0.00 | 2.05 | 1.05 | 0.00 | 1.23 | 1.01 | 0.07 | 1.23 | 1 | 1.00 | 1.23 | 1.04 | 0.00 | 1.32 |
| | MONK_S | 0.94 | 0.00 | 1.50 | 1 | 0.73 | 5.04 | 1.13 | 0.00 | 2.24 | 1.05 | 0.00 | 2.21 | 1.02 | 0.01 | 1.53 | 1.01 | 0.16 | 1.53 | 1.05 | 0.00 | 1.80 |
| | MONK_L | 0.91 | 0.00 | 2.49 | 0.41 | 0.00 | 3.60 | 0.89 | 0.00 | 1.62 | 0.93 | 0.00 | 2.15 | 0.94 | 0.00 | 1.79 | 0.95 | 0.01 | 2.32 | 0.8 | 0.00 | 1.57 |
| | HMONK_S_C0 | 0.95 | 0.00 | 2.22 | 0.96 | 0.85 | 5.31 | 1.11 | 0.00 | 2.22 | 1.03 | 0.02 | 2.65 | 1 | 0.48 | 2.38 | 0.98 | 0.00 | 2.38 | 1.04 | 0.01 | 2.88 |
| | HMONK_S_C1 | 0.99 | 0.08 | 1.79 | 0.99 | 0.03 | 2.87 | 1.11 | 0.00 | 1.92 | 1.08 | 0.00 | 1.07 | 1.05 | 0.00 | 1.07 | 1.03 | 0.00 | 1.07 | 1.06 | 0.00 | 1.18 |
| | HMONK_S_C2 | 0.99 | 0.00 | 0.47 | 0.96 | 0.10 | 2.78 | 1.09 | 0.00 | 1.82 | 1.08 | 0.00 | 1.49 | 1.07 | 0.00 | 1.31 | 1.06 | 0.00 | 1.04 | 1.06 | 0.00 | 1.28 |
| | HMONK_S_C3 | 0.99 | 0.93 | 1.35 | 1.01 | 0.43 | 3.05 | 1.07 | 0.00 | 2.99 | 1.08 | 0.00 | 1.14 | 1.08 | 0.00 | 0.90 | 1.06 | 0.00 | 1.75 | 1.07 | 0.00 | 1.44 |
| | HMONK_S_C4 | 0.99 | 0.11 | 1.38 | 1.03 | 0.78 | 7.33 | 1.04 | 0.00 | 1.82 | 1.06 | 0.00 | 2.29 | 1.06 | 0.00 | 2.41 | 1.06 | 0.00 | 2.08 | 1.05 | 0.00 | 2.64 |
| | HMONK_C0 | 0.98 | 0.00 | 1.02 | 1.03 | 0.07 | 3.55 | 1.15 | 0.00 | 0.73 | 1.08 | 0.00 | 1.01 | 1.04 | 0.00 | 1.01 | 1.03 | 0.00 | 1.07 | 1.06 | 0.00 | 1.29 |
| | HMONK_C1 | 0.99 | 0.14 | 1.01 | 1.02 | 0.50 | 4.57 | 1.12 | 0.00 | 1.10 | 1.08 | 0.00 | 1.70 | 1.05 | 0.00 | 1.46 | 1.03 | 0.00 | 1.39 | 1.06 | 0.00 | 1.48 |
| | HMONK_C2 | 0.98 | 0.02 | 1.50 | 1.01 | 0.36 | 2.16 | 1.08 | 0.00 | 2.51 | 1.07 | 0.00 | 1.64 | 1.06 | 0.00 | 1.05 | 1.05 | 0.00 | 1.31 | 1.05 | 0.00 | 1.32 |
| | HMONK_C3 | 0.99 | 0.00 | 0.64 | 1.03 | 0.17 | 4.13 | 1.07 | 0.00 | 1.05 | 1.07 | 0.00 | 1.07 | 1.07 | 0.00 | 0.52 | 1.06 | 0.00 | 0.91 | 1.06 | 0.00 | 0.75 |
| | HMONK_C4 | 0.99 | 0.22 | 1.29 | 1.04 | 0.02 | 4.63 | 1.06 | 0.00 | 1.65 | 1.07 | 0.00 | 1.11 | 1.08 | 0.00 | 1.44 | 1.07 | 0.00 | 0.99 | 1.07 | 0.00 | 0.77 |
| AMD | vanilla | 1 | – | 0.54 | 1 | – | 1.89 | 1 | – | 0.63 | 1 | – | 1.24 | 1 | – | 1.08 | 1 | – | 1.69 | 1 | – | 0.81 |
| | MONK | 0.98 | 0.00 | 0.55 | 1.03 | 0.01 | 1.30 | 1.08 | 0.00 | 1.10 | 1.04 | 0.00 | 0.92 | 1.02 | 0.01 | 0.71 | 1.0 | 0.75 | 0.58 | 1.03 | 0.00 | 0.52 |
| | MONK_S | 0.98 | 0.00 | 0.67 | 1.01 | 0.16 | 2.06 | 1.08 | 0.00 | 0.58 | 1.04 | 0.00 | 0.71 | 1.02 | 0.07 | 0.71 | 1.0 | 0.24 | 1.16 | 1.03 | 0.00 | 0.61 |
| | MONK_L | 0.98 | 0.11 | 1.59 | 0.86 | 0.00 | 1.03 | 0.95 | 0.00 | 0.66 | 0.95 | 0.00 | 1.09 | 0.94 | 0.00 | 0.66 | 0.93 | 0.00 | 0.76 | 0.93 | 0.00 | 0.61 |
| | HMONK_S_C0 | 0.98 | 0.00 | 0.67 | 1.02 | 0.32 | 4.40 | 1.08 | 0.00 | 0.93 | 1.04 | 0.00 | 1.16 | 1.02 | 0.01 | 1.08 | 1.01 | 0.46 | 1.46 | 1.03 | 0.00 | 1.30 |
| | HMONK_S_C1 | 0.98 | 0.00 | 1.34 | 1.02 | 0.00 | 1.24 | 1.08 | 0.00 | 1.44 | 1.06 | 0.00 | 0.76 | 1.04 | 0.00 | 0.57 | 1.01 | 0.11 | 0.57 | 1.04 | 0.00 | 0.67 |
| | HMONK_S_C2 | 0.99 | 0.00 | 0.86 | 1.01 | 0.71 | 2.46 | 1.09 | 0.00 | 0.56 | 1.06 | 0.00 | 1.15 | 1.04 | 0.00 | 1.66 | 1.01 | 0.13 | 1.66 | 1.04 | 0.00 | 0.86 |
| | HMONK_S_C3 | 0.99 | 0.00 | 0.86 | 1.01 | 0.95 | 1.57 | 1.08 | 0.00 | 0.58 | 1.07 | 0.01 | 0.66 | 1.05 | 0.00 | 0.55 | 1.03 | 0.11 | 0.09 | 1.05 | 0.01 | 0.36 |
| | HMONK_S_C4 | 0.99 | 0.09 | 1.22 | 1.00 | 0.61 | 3.21 | 1.08 | 0.00 | 1.60 | 1.07 | 0.00 | 1.44 | 1.06 | 0.00 | 0.68 | 1.04 | 0.01 | 0.68 | 1.05 | 0.00 | 0.83 |
| | HMONK_Co* | 0.98 | 0.00 | 0.62 | 1.0 | 0.79 | 3.85 | 1.08 | 0.00 | 0.66 | 1.04 | 0.04 | 1.07 | 1.01 | 0.27 | 1.07 | 1.0 | 0.75 | 0.75 | 1.02 | 0.01 | 1.17 |
| | HMONK_C1 | 0.98 | 0.00 | 1.10 | 0.99 | 0.17 | 4.05 | 1.09 | 0.00 | 0.71 | 1.04 | 0.00 | 0.71 | 1.02 | 0.02 | 0.71 | 1.0 | 0.65 | 1.82 | 1.03 | 0.01 | 1.41 |
| | HMONK_C2 | 0.99 | 0.00 | 0.86 | 1.01 | 0.28 | 2.56 | 1.10 | 0.00 | 1.08 | 1.05 | 0.00 | 1.01 | 1.04 | 0.00 | 0.90 | 1.02 | 0.02 | 1.17 | 1.04 | 0.00 | 0.80 |
| | HMONK_C3 | 0.99 | 0.00 | 0.54 | 0.99 | 0.95 | 2.54 | 1.09 | 0.00 | 1.48 | 1.07 | 0.01 | 1.39 | 1.05 | 0.00 | 1.13 | 1.03 | 0.11 | 1.38 | 1.04 | 0.01 | 1.14 |
| | HMONK_C4 | 0.99 | 0.00 | 0.86 | 1.0 | 0.10 | 2.04 | 1.09 | 0.00 | 1.16 | 1.07 | 0.00 | 0.69 | 1.05 | 0.00 | 0.89 | 1.03 | 0.02 | 0.89 | 1.05 | 0.00 | 0.58 |

# Monk: Opportunistic Scheduling to Delay Horizontal Scaling

**Table 3** Aggregated latency of 30 runs of Hazelcast on the Intel machine normalized to unmodified ZGC – lower is better. Green indicates statistically better, red indicates statistically worse, and white shows the results without statistical significance. p-value is the result of the Welch test. RSD is the relative standard deviation in %.

| Folder | BM | val | p-val | RSD | BM | val | p-val | RSD | BM | val | p-val | RSD |
|---|---|---|---|---|---|---|---|---|---|---|---|---|
| Vanilla |        | 1    | –    | 29.57 |        | 1    | –    | 18.04 |        | 1    | –    | 18.79 |
| MONK    | hz_80  | 0.62 | 0.00 | 25    | hz_190 | 0.61 | 0.00 | 12.19 | hz_290 | 1.69 | 0.01 | 77.03 |
| MONK_S  |        | 0.56 | 0.00 | 28.05 |        | 0.64 | 0.00 | 11.69 |        | 1.88 | 0.00 | 76.52 |
| Vanilla |        | 1    | –    | 15.08 |        | 1    | –    | 17.18 |        | 1    | –    | 23.69 |
| MONK    | hz_100 | 0.57 | 0.00 | 19.82 | hz_200 | 0.67 | 0.00 | 15.39 | hz_300 | 2.38 | 0.00 | 49.74 |
| MONK_S  |        | 0.59 | 0.00 | 20.91 |        | 0.62 | 0.00 | 0     |        | 2.19 | 0.00 | 68.04 |
| Vanilla |        | 1    | –    | 22.63 |        | 1    | –    | 18.34 |        | 1    | –    | 34.27 |
| MONK    | hz_110 | 0.59 | 0.00 | 20.8  | hz_210 | 0.65 | 0.00 | 15.3  | hz_315 | 2.72 | 0.00 | 46.11 |
| MONK_S  |        | 0.62 | 0.00 | 23.88 |        | 0.62 | 0.00 | 15.23 |        | 3.6  | 0.00 | 61.5  |
| Vanilla |        | 1    | –    | 15.04 |        | 1    | –    | 12.74 |        | 1    | –    | 31.85 |
| MONK    | hz_120 | 0.58 | 0.00 | 16.26 | hz_220 | 0.66 | 0.00 | 14.71 | hz_325 | 3.91 | 0.00 | 38.1  |
| MONK_S  |        | 0.56 | 0.00 | 19.19 |        | 0.66 | 0.00 | 18.63 |        | 2.34 | 0.00 | 31    |
| Vanilla |        | 1    | –    | 16.58 |        | 1    | –    | 10.41 |        | 1    | –    | 88.87 |
| MONK    | hz_130 | 0.65 | 0.00 | 19.45 | hz_230 | 0.59 | 0.00 | 10.37 | hz_335 | 1.17 | 0.45 | 66.42 |
| MONK_S  |        | 0.62 | 0.00 | 18.21 |        | 0.61 | 0.00 | 13.69 |        | 0.68 | 0.12 | 82.93 |
| Vanilla |        | 1    | –    | 13.5  |        | 1    | –    | 14.55 |        | 1    | –    | 64.12 |
| MONK    | hz_140 | 0.67 | 0.00 | 18.66 | hz_240 | 0.59 | 0.00 | 13.25 | hz_350 | 1.29 | 0.20 | 71.32 |
| MONK_S  |        | 0.63 | 0.00 | 18.43 |        | 0.61 | 0.00 | 10.88 |        | 1.24 | 0.20 | 51.4  |
| Vanilla |        | 1    | –    | 9.65  |        | 1    | –    | 15.5  |        | 1    | –    | 38.36 |
| MONK    | hz_150 | 0.69 | 0.00 | 16.77 | hz_250 | 0.64 | 0.00 | 9.28  | hz_360 | 1.05 | 0.68 | 51.76 |
| MONK_S  |        | 0.65 | 0.00 | 19.45 |        | 0.63 | 0.00 | 12.09 |        | 0.95 | 0.76 | 66.81 |
| Vanilla |        | 1    | –    | 16.35 |        | 1    | –    | 12.61 |        | 1    | –    | 42.73 |
| MONK    | hz_160 | 0.63 | 0.00 | 17.35 | hz_260 | 0.62 | 0.00 | 12.07 | hz_375 | 0.28 | 0.00 | 38.87 |
| MONK_S  |        | 0.62 | 0.00 | 17.98 |        | 0.58 | 0.00 | 10.22 |        | 0.3  | 0.00 | 43.4  |
| Vanilla |        | 1    | –    | 13.78 |        | 1    | –    | 14.82 |        | 1    | –    | 19.13 |
| MONK    | hz_170 | 0.72 | 0.00 | 19.07 | hz_270 | 1.04 | 0.75 | 59.99 | hz_400 | 0.82 | 0.00 | 27.32 |
| MONK_S  |        | 0.65 | 0.00 | 20.02 |        | 0.69 | 0.00 | 24.7  |        | 0.79 | 0.00 | 16.75 |
| Vanilla |        | 1    | –    | 13.5  |        | 1    | –    | 11.91 |        | 1    | –    | 4.64  |
| MONK    | hz_180 | 0.73 | 0.00 | 14.72 | hz_280 | 1.73 | 0.01 | 74.86 | hz_500 | 0.95 | 0.01 | 8.28  |
| MONK_S  |        | 0.75 | 0.00 | 18.49 |        | 0.57 | 0.00 | 19.02 |        | 0.93 | 0.00 | 7.33  |

with a primary focus on reducing latency. Achieving this goal entails considering several crucial aspects: (1) limiting the maximum pause time introduced by garbage collection, (2) minimizing the frequency with which a mutator can be preempted or paused by garbage collection, (3) addressing the potential worst-case slowdown resulting from write barriers that accompany garbage collection. To tackle these challenges, the paper explores two distinct scheduling strategies: slack-based scheduling and periodic scheduling. Slack-based scheduling involves running the garbage collector at a lower priority, which helps alleviate pause time but does not directly address the worst-case slowdown caused by write barriers. Periodic scheduling, on the other hand, prioritizes the garbage collector to preempt mutators at regular intervals, allowing it to perform a predefined amount of collector work within a time-bound window. However, this strategy also does not guarantee a solution to the worst-case slowdown problem.

Ulan Degenbaev et al. [10] present an implementation of slack-based scheduling within the V8 JavaScript engine in Chrome, specifically within the Blink scheduler. This approach differs slightly from previous slack-based scheduling methods. Here, the terms slack or idle time pertains to those moments when Chrome has no other tasks to execute. The original scheduler in Chrome/Blink employs multiple task queues, where tasks within the same queue must execute sequentially, while tasks between different queues can be reordered. In the context of idle tasks—tasks to be executed when there are no other pending tasks—an additional queue is introduced, allocated with the lowest priority. A notable feature of this scheduling strategy is that it never aborts garbage collection tasks; instead, V8 schedules them meticulously to ensure they fit within available idle time intervals. This approach optimizes resource utilization and ensures that garbage collection tasks are efficiently executed while adhering to the specific needs of the Chrome browser and V8 engine.

Junxian Zhao et al. [36] proposed a middleware that adaptively triggers GC based on the CPU utilization at runtime to reduce latency. Our approach does not require additional software and further investigation is required to compare the performance of our approach for multi-tenant environments.

In conclusion, our study builds upon the foundation laid by previous research on slack-based GC scheduling. The techniques employed in our study are derived from the aforementioned works. However, we extend this well-established approach to a novel environment—server workloads—with a distinct objective: delaying horizontal scaling.

By venturing into this new domain, we have the opportunity to investigate a simpler solution tailored to our specific needs. This simplicity offers easier integration and maintenance within a production GC environment.

# 7 Conclusion

The overarching goal of our work is to facilitate lowering the carbon footprint of ICT. We believe that Monk can contribute to this by extending the range of CPU utilization under which an application can meet tight SLAs. This permits delaying the point in time before scaling out, which leads to hardware utilization improvement.

**Monk: Opportunistic Scheduling to Delay Horizontal Scaling**

Our immediate goal with Monk has been to explore the potential of a very simple idea: down-prioritize GC threads so that GC only runs when there are idle resources. Our results show that such a design can improve response times on moderate CPU loads. This improvement comes from reducing the competition for CPU resources between GC threads and mutators. We explored multiple approaches to dealing with priority inversion causing GC performance to affect mutator performance directly: by reverting to the default scheduling policy during stop-the-world pauses, and when holding locks. We believe that in reasonable deployment scenarios with enough resource headroom, the benefit of priority inversion inside locks and safepoints does not outweigh the cost.

Finally, we explored how to handle GC starvation, which invariably happens under high CPU load in our design. We found that addressing the starvation of GC processes is crucial for application performance. This is especially true for the baseline we chose for our prototype implementation, as it only briefly stops applications inside stop-the-world pauses and locks. We believe that a fallback strategy can significantly reduce the negative effects of GC starvation on high CPU loads, although work is needed to choose the suitable threshold for backing off so as not to back off too early to negate latency benefits.

**Acknowledgements**   This research was supported by the Swedish Research Council through the project Accelerating Managed Languages (2020-05346), by the Swedish Foundation for Strategic Research through the project Deploying Memory Management Research in the Mainstream (SM19-0059), and by donations from Oracle Corporation.

## A  High MONK Retaining Vanilla Behavior During Critical GC Phases

**Table 4** Aggregated performance metrics of 5 runs of SPECjbb2015 normalized to unmodified ZGC – higher is better. Green indicates statistically better, red indicates statistically worse, and white shows the results without statistical significance. p-value is the result of the Welch test. RSD is the relative standard deviation in %.

| HW | Conf | max-JOPS | | | 10 ms | | | 25 ms | | | 50 ms | | | 75 ms | | | 100 ms | | | critical-JOPS | | |
|---|---|---|---|---|---|---|---|---|---|---|---|---|---|---|---|---|---|---|---|---|---|---|
| | | val | p-val | RSD | val | p-val | RSD | val | p-val | RSD | val | p-val | RSD | val | p-val | RSD | val | p-val | RSD | val | p-val | RSD |
| Intel | HMONK_C0 | 1.01 | 0.72 | 2.11 | 1.00 | 0.87 | 3.90 | 0.99 | 0.46 | 0.77 | 0.98 | 0.39 | 1.72 | 0.98 | 0.13 | 1.20 | 0.99 | 0.40 | 1.64 | 0.98 | 0.35 | 1.55 |
| | HMONK_C1 | 1.00 | 0.93 | 2.03 | 1.02 | 0.27 | 2.16 | 0.96 | 0.09 | 1.65 | 0.99 | 0.79 | 1.73 | 0.99 | 0.47 | 1.94 | 0.99 | 0.58 | 1.42 | 0.99 | 0.55 | 1.44 |
| | HMONK_C2 | 0.99 | 0.43 | 1.84 | 1.00 | 0.89 | 3.64 | 1.10 | 0.01 | 1.03 | 1.09 | 0.01 | 1.74 | 1.05 | 0.02 | 1.68 | 1.03 | 0.14 | 1.49 | 1.05 | 0.02 | 1.49 |
| | HMONK_C3 | 0.99 | 0.50 | 2.51 | 1.00 | 0.98 | 2.13 | 1.04 | 0.17 | 4.35 | 1.03 | 0.20 | 2.33 | 1.01 | 0.34 | 0.68 | 1.00 | 0.80 | 1.02 | 1.02 | 0.29 | 1.58 |
| | HMONK_C4 | 1.00 | 0.85 | 3.23 | 1.01 | 0.43 | 1.75 | 1.02 | 0.65 | 5.96 | 1.01 | 0.70 | 3.33 | 1.00 | 0.82 | 2.08 | 0.99 | 0.58 | 1.11 | 1.01 | 0.77 | 2.68 |

The first observation from Table 4 is that the latency data is slightly less stable compared to the "every-second-cycle" High MONK configuration, presented in Table 2. This instability arises because, in this version, we allow greater contention between mutators and GC threads, which introduces more variability in latency.

Secondly, this version matches Vanilla's throughput, whereas the "every-second-cycle" High MONK, as discussed in Section 5, experiences a minor throughput regression. This outcome aligns with our expectations since reverting to Vanilla behavior during critical GC phases should result in a similar performance to Vanilla, with no noticeable throughput differences.

Although this version captures the full potential in terms of throughput, the latency improvement is more modest. Only the HMONK_C2 variant achieves statistically significant improvements in critical-jops, similar to the "every-second-cycle" HMONK_C2. The other variants show some latency improvements, but they are not statistically significant, likely due to greater variability in the data.

In conclusion, when designing a production solution, it is crucial not only to address priority inversion but also to consider the frequency of priority switching carefully. This balance introduces a trade-off between throughput and latency, both of which must be optimized based on the system's specific requirements.

## About the authors

**Marina Shimchenko** is a recently graduated PhD student at Uppsala University. Contact her at marina.shimchenko@it.uu.se.
https://orcid.org/0000-0002-0701-8540

**Erik Österlund** is a Principal Engineer at Oracle in Sweden. Contact him at erik.osterlund@oracle.com.
https://orcid.org/0000-0003-3686-8568

**Tobias Wrigstad** is a professor in computing science at Uppsala University. Contact him at tobias.wrigstad@it.uu.se.
https://orcid.org/0000-0002-4269-5408